# PixelsDB: Serverless and NL-Aided Data Analytics with Flexible Service Levels and Prices

Haoqiong Bian, Dongyang Geng, Haoyang Li, Yunpeng Chai
Renmin University of China
{bianhq,gengdongyang,lihaoyang.cs,ypchai}@ruc.edu.cn

Anastasia Ailamaki
EPFL
anastasia.ailamaki@epfl.ch

*Abstract*—Serverless query processing has become increasingly popular due to its advantages, including automated resource management, high elasticity, and pay-as-you-go pricing. For users who are not system experts, serverless query processing greatly reduces the cost of owning a data analytic system. However, it is still a significant challenge for non-expert users to transform their complex and evolving data analytic needs into proper SQL queries and select a serverless query service that delivers satisfactory performance and price for each type of query.

This paper presents PixelsDB, an open-source data analytic system that allows users who lack system or SQL expertise to explore data efficiently. It allows users to generate and debug SQL queries using a natural language interface powered by fine-tuned language models. The queries are then executed by a serverless query engine that offers varying prices for different performance service levels (SLAs). The performance SLAs are natively supported by dedicated architecture design and heterogeneous resource scheduling that can apply cost-efficient resources to process non-urgent queries. We demonstrate that the combination of a serverless paradigm, a natural-language-aided interface, and flexible SLAs and prices will substantially improve the usability of cloud data analytic systems.

## I. INTRODUCTION

Serverless query processing, also known as Query-as-a-Service (QaaS), has become the new paradigm of analytical query processing in the cloud. In contrast to traditional 'serverful' query engines, serverless query services like AWS Athena [1] and BigQuery [2] automate laborious tasks such as resource provisioning and cluster scaling. This allows users to run queries without the burden of resource management. Moreover, users are billed on a per-query basis according to the number of processing units used (*e.g.*, Redshift-serverless [3]) or amount of data scanned (*e.g.*, AWS Athena [1] and BigQuery [2]) by the queries. In a word, for users who are not system experts, serverless query processing greatly reduces the effort required to own and use a data analytic system.

However, through a user study among database practitioners, we find that users of serverless query services have further demands for flexible performance-prices trade-offs and natural-language-aided query interfaces [4]. In the user study, we sent questionnaires to 887 database practitioners through Tencent Survey [5] and got 109 valid submissions[1]. Of the valid submissions, 100 prefer serverless query processing (with automated on-demand resource allocation) over the traditional resource-provisioned query engines. Furthermore, as shown in Figure 1, among the users who prefer serverless query processing, 79% prefer to have the flexibility of choosing a specific service level (of performance and price) for each query, while 84% users would like to try or use a natural-language-aided query interface. This actually aligns with our intuition. Imagine we are buying products in a shop. We would hope that each product has a label with its performance and price, and the product can be handled in a simple way. However, these demands are not addressed well in existing serverless query services. Redshift-serverless provides users with a bar to adjust performance-cost trade-offs [6]. However, the adjustment applies to the entire cluster rather than an individual query, and it is not effective immediately as it requires collecting a long-term query history.

[1]The questionnaire, validation rules, and valid and invalid submissions are available at https://github.com/pixelsdb/cdw-user-study

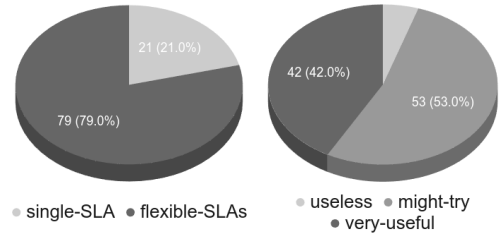

(a) Service levels.    (b) NL-aided interface.

Fig. 1: Preferences of serverless query service users on service levels and the natural-language-aided query interface.

This paper presents an open-source serverless query system named *PixelsDB* (https://github.com/pixelsdb/pixels). It combines text-to-SQL and flexible SLAs to address the aforementioned user demands, thus improving the usability of serverless query services for non-expert users. One major intellectual contribution in this paper is to propose an approach to natively supporting varying service levels (in terms of query urgency, say *immediate*, *relaxed*, and *best-of-effort*) and prices through dedicated architecture design and heterogeneous resource scheduling [4]. This conforms to users' natural classification of interactive (*e.g.*, ad-hoc and busy-dashboard queries) and non-interactive queries (*e.g.*, data-report and off-peak queries). For non-interactive queries, choosing a lower service level (*i.e.*, relaxed or best-of-effort) allows the system to schedule cost-efficient resources for query execution, thereby reducing query costs (prices) for users.

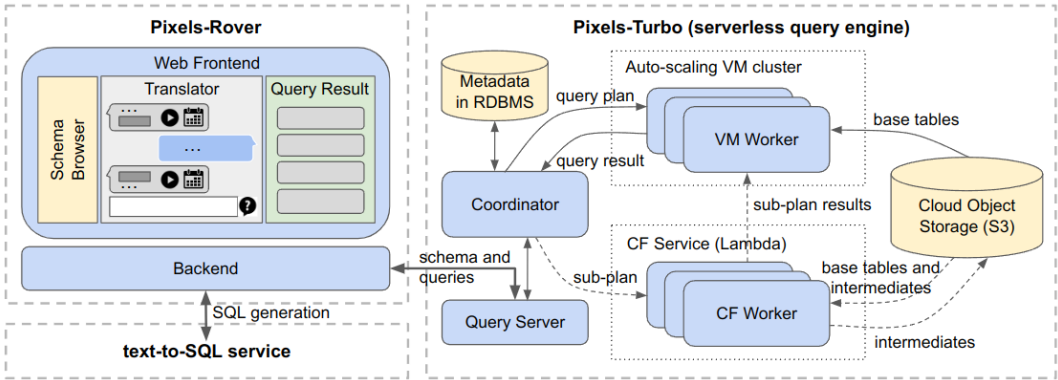

Fig. 2: Overview of the user interface and system architecture of PixelsDB.

## II. SYSTEM OVERVIEW

As shown in Figure 2, PixelsDB is composed of three parts.

**(1) Pixels-Rover** provides the user interface demonstrated in this paper. It is the frontend that connects to the text-to-SQL service and the serverless query engine. The user first logs in to Pixels-Rover, views the authorized database schemas, and selects a database to analyze. Then, the user's analytic questions typed in the *Translator* are sent to the text-to-SQL service for SQL translation. Each query translated by the text-to-SQL service is displayed as a code block below the user's question. The user can edit the query in the code block and submit it with a desired service level (and corresponding price) to the serverless query engine. Pixels-Rover provides three service levels: (a) Immediate, which starts to execute the query immediately; (b) Relaxed, which starts to execute the query in a configurable grace period; and (c) Best-of-effort, which executes the query at best effort with no guarantee. The status and result of each submitted query are shown in the *Query Result* area and are linked to the origin query in the translator.

**(2) Pixels-Turbo** is responsible for serverless query processing and providing flexible service levels and prices. The architecture of *Pixels-Turbo* is shown in Figure 2. The solid and dotted arrows are the required and optional steps to process a query. The *Query Server* provides a REST API to receive queries from clients (*e.g.*, Pixels-Rover) and interacts with the *Coordinator* to fetch database schema or schedule the queries to execute at specified service levels. The coordinator runs in a small VM, which is the only long-running component in the query engine. It is responsible for managing metadata, parsing queries, allocating resources, coordinating query execution tasks, and collecting the query results and statistics (*e.g.*, execution time and resource consumption).

The *VM cluster* and the *CF Service* are the two computing resources that Pixels-Turbo adaptively uses to execute queries. The VM cluster is more cost-efficient for processing stable workloads but requires 1-2 minutes to scale for workload changes. Whereas the CF service is more elastic (*e.g.*, create hundreds of workers in 1 second) but has 9-24x higher resource unit prices [7]. Base tables and CF-produced intermediate results are stored in cloud object storage, such as AWS S3. The solution for supporting various service levels based on heterogeneous resource scheduling is discussed in [4] and briefly introduced in Section III-A.

**(3) Text-to-SQL service** provides high-quality text-to-SQL translation. Whenever the user submits a natural language question, Pixels-Rover compiles a message containing the question and the schema elements (*e.g.*, table and column names) of the user's selected database and sends it to the text-to-SQL service. Then, the service translates the question into an SQL query and responds to Pixels-Rover.

Note that the text-to-SQL service in PixelsDB is plug-able. Thus, we can choose any high-quality text-to-SQL services. We designed a unified wrapper interface for text-to-SQL services in Pixels-Rover. It allows Pixels-Rover to access a text-to-SQL service by plugging in the corresponding wrapper. Currently, we use CodeS [8] as the text-to-SQL service in the on-premises deployment of PixelsDB. CodeS shows state-of-the-art performance on challenging text-to-SQL benchmarks such as Spider [9] and BIRD [10]. We can support other text-to-SQL services by implementing new wrappers. Optimizing and benchmarking different text-to-SQL techniques is beyond the scope of this demonstration.

## III. KEY TECHNOLOGIES

In this section, we introduce the key technologies behind PixelsDB, including how queries are executed on heterogeneous computing infrastructures with varying elasticity and cost-efficiency (Section III-A) and how flexible service levels and prices are implemented on the aforementioned query runtime (Section III-B).

### A. Query Execution

Virtual machine (VM) and cloud function (CF) are the two major types of computing services in the cloud. As discussed in Section II, VM is more cost-efficient but less elastic than CF. Therefore, they can be complementary in supporting elastic and cost-efficient query processing. In *Pixels-Turbo*, the *Coordinator* can dynamically create CF Workers to execute the new coming queries when the VM cluster does not have enough resources and can not scale out in time. This is done by pushing down the expensive operators (*e.g.*, table scans, joins, and aggregations) from the top-level plan of the new coming query into a sub-plan. The ephemeral CF workers are then launched to execute the sub-plan and return its result to the top-level plan running in the VM cluster. Thus, the query is executed without further overloading the VM cluster, and this is transparent to the query clients. After 1-2 minutes, when the VM cluster scales to an appropriate capacity, CF will no longer participate in processing the coming queries. More details about query execution in CFs are discussed in [7].

As shown in Figure 2, the VM cluster is auto-scaled following workload changes (although it may have 1-2 minutes of lag). Such auto-scaling actions are triggered by the scaling manager and metrics collector in the Coordinator. The metrics collector collects performance metrics, such as query concurrency and CPU/memory load, from the VM cluster. The scaling manager monitors the performance metrics and runs a scaling policy to decide whether to release or create VMs for the VM cluster. The scaling policy is plug-able and configurable.

To support different performance-price trade-offs for the service levels, when the query client is submitting a query, we allow it to indicate (1) whether a pending time is acceptable for query execution and (2) whether CF-based acceleration is acceptable for query execution. If neither is acceptable, Pixels-Turbo may invoke high-elastic resources (e.g., CFs) for acceleration if the cost-efficient resources (e.g., VMs) are not enough. If both are acceptable, Pixels-Turbo will wait for at most N minutes (configurable) for the cost-efficient resources to be available and will invoke high-elastic resources if the grace period expires. This increases the opportunities for cost savings and provides a pending time guarantee. If only (1) is acceptable, Pixels-Turbo will never execute the query using high-elastic resources. The query, in this case, has the lowest priority and is only executed when the cost-efficient resources are idle. Thus, the query will not trigger the scaling-out action of the cost-efficient resources, hence producing lower costs. These three cases correspond to the three service levels *Immediate*, *Relaxed*, and *Best-of-effort* discussed in Section III-B. More details about query scheduling in Pixels-Turbo are discussed in [4].

### B. Flexible Service Levels and Prices

The *Query Server* of Pixels-Turbo provides three service levels and corresponding prices for each query submission.

**(1) Immediate.** At this service level, the query server receives the query from the client and immediately submits it to the coordinator with CF acceleration enabled. Thus, this service level guarantees immediate execution but a higher price upper bound, as the expensive CFs might be involved in query execution. In this demo, we set the price of immediate queries to be the same as AWS Athena [2], which is $5/TB-scan. We also log the actual resource costs of each query in the backend.

**(2) Relaxed.** At this service level, the query can be queued in the query server before a configurable grace period (*e.g.*, 5 minutes) expires, giving time for the VM cluster to scale out. Given a grace period longer than the time required to scale out the VM cluster, relaxed queries will not overload the VM cluster. Evaluations show that under continuous workload, this service level generally produces 2-5x lower resource costs than Immediate. Hence, in this demo, we set the price of relaxed queries to 40% of the immediate queries, *i.e.*, $2/TB-scan.

**(3) Best-of-effort.** At this service level, the query server only schedules the query for execution when the VM cluster is idle. There is no guarantee on the pending time. Namely, best-of-effort queries are only executed when the VM cluster is likely to scale in. This helps the VM cluster avoid unnecessary scaling-in [3] and produces very little extra costs. Evaluations show that under continuous workload, this service level generally produces more than one order of magnitude lower resource costs than immediate. Hence, in this demo, we set the price of best-of-effort queries to 10% of the immediate queries, *i.e.*, $0.5/TB-scan.

Note that each service level only specifies the upper bound on query pending time. Relaxed or best-of-effort queries may be executed immediately if the VM cluster is available.

### IV. DEMONSTRATION

In this section, we demonstrate how users can interact with PixelsDB in two typical use cases.

#### A. Use Case 1: Interactive Analytics

After logging in, we can see the main user interface of Pixels-Rover shown in Figure 3. It is composed of three components: a left sidebar containing the schema browser and historical query statistics reporter, a translator panel for submitting questions and queries, and a query result panel for checking the query results. We can interact with these components to do data analysis in three main steps.

*1) Query translation:* We first select the database to analyze in the drop-down box at the lower left of the *Translator*. Then, we type in our question in the message box and click *Send*. Each question is translated to an SQL query and displayed as a code block below the question. When we hover over the SQL code block, as shown by the last query in Figure 3, we can see a dynamic edit button ✏ and a dynamic submit button ➤ on the right of the code block. If we want to correct minor errors in the query, we can make the code block writable by clicking the edit button. After editing the query, we can click the dynamic cancel button ✖ on the right of the code block to reset the query or click the dynamic confirm button ✓ to accept the modification.

*2) Submit query with a preferred service level:* When we are satisfied with the query, we can click the submit button, and a submission form (shown in Figure 4) will pop up as a translucent floating layer on the web page. In this form, we can select the service level and set the result-size limit for the query. By clicking the submit button ➤ on this form, we submit the query to Pixels-Turbo, and the query will be scheduled for execution.

---

[2]In the experiments of [7], we show that the pure-CF query execution in Pixels-Turbo has a comparable monetary cost as AWS Athena and Redshift-Serverless.

[3]For example, scaling-in right before the next workload spike. We tried to avoid this by a lazy-scaling-in policy in [7].

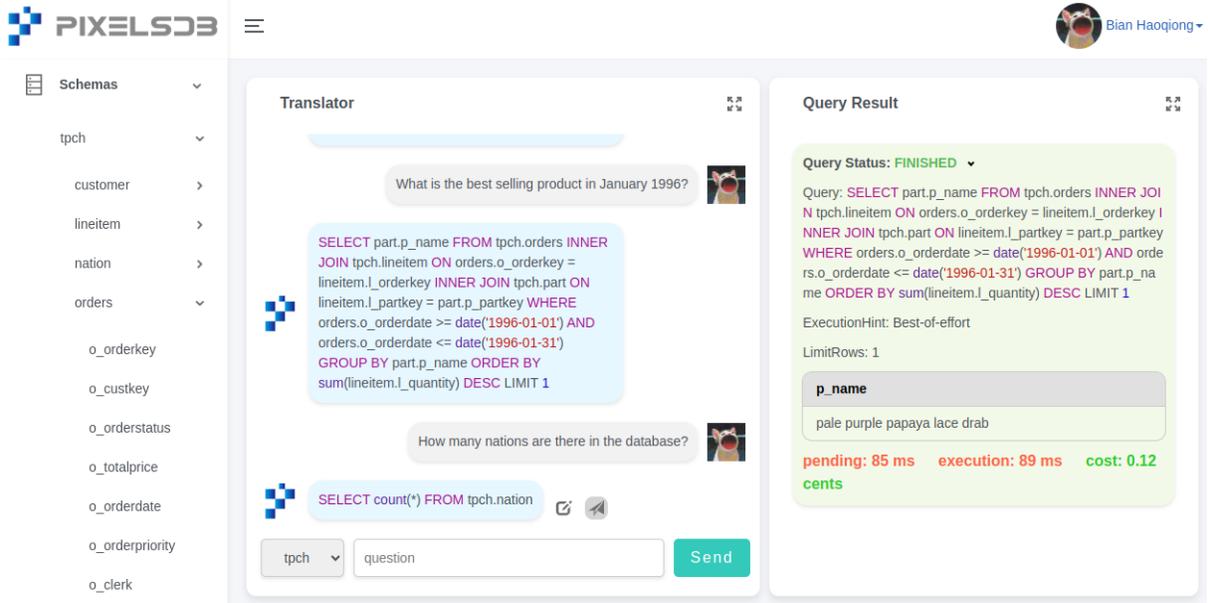

Fig. 3: Main user interface of PixelsDB (provided by Pixels-Rover).

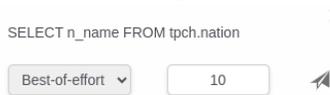

Fig. 4: The submission form of a query.

*3) Check query status and result:* After the query is submitted, an expendable block that displays the query's status and result will appear in the Query Result area. All blocks are arranged in ascending order from top to bottom according to query submission time. We use the background colors ▮, ▮, and ▮ for the blocks of best-of-effort, relaxed, and immediate queries, respectively. We can double-click a query's code block in the translator to highlight its status-and-result block, or vice versa. There are four query statuses: pending (*i.e.*, waiting to execute), running (*i.e.*, executing), finished, and failed. If the query is finished, we can expand the block to view its result and execution statistics (*i.e.*, pending time, execution time, and monetary cost). If the query is failed, we can expand its block to view the error message. To expand or restore the corresponding area, we can click the zoom button at the upper right of the translator or query result.

### B. Analysis Query Statistics

After a long session of interactive analytics, we may want to analyze the performance, costs, and service levels of the historical queries, just like checking the monthly credit card bills. We call this function *cost visibility*. The user study mentioned in Section 1 also shows that database users have a strong interest in this function.

This is supported by the *Report* tab in the left sidebar, just below the schema browser. By clicking the Report tab, we can see three charts reporting the query count per minute in the timeline, query performance (pending time and execution time) of each query, and query cost of each query, respectively. The query performance chart and the query cost chart are brush-and-linked to the query count timeline chart. The performance and costs of the selected queries will be shown in the other two charts by brushing a segment using the mouse on the timeline. By clicking a query in the performance or cost chart, we can also see the detailed query information in the query result panel.

## V. CONCLUSION

This paper presents a data analytic system named *PixelsDB*. It demonstrates how serverless query processing, text-to-SQL translation, and flexible service levels and prices can be integrated to improve the user experience of cloud data warehouses (lakes). The different levels of query pending time in PixelsDB conform to users' natural classification of interactive and non-interactive queries. This allows the system to schedule resources with different levels of elasticity and cost efficiency to execute different types of queries. It also provides opportunities for batch query optimization.